\documentclass[12pt]{article}
\usepackage{epsfig}
\usepackage{rotate}
\usepackage{psfrag}
\usepackage{a4}

\newcommand{\be}{\begin{equation}}
\newcommand{\ee}{\end{equation}}
\newcommand{\beqas}{\begin{eqnarray*}}
\newcommand{\eeqas}{\end{eqnarray*}}
\newcommand{\beqar}{\begin{eqnarray}}
\newcommand{\eeqar}{\end{eqnarray}}

\begin{document}
\title{Correlation structure of extreme stock returns}

\author{
Pierre Cizeau$^\dagger$,
Marc Potters$^\dagger$ and
Jean-Philippe Bouchaud$^{*,\dagger}$}
\vspace{0.5cm}
\date{\sl \small
$^\dagger$Science \& Finance\\
The Research Division of Capital Fund Management\\
109--111 rue Victor-Hugo, 92532 Levallois {\sc cedex}, France \\
http://www.science-finance.fr \\
\vspace{0.4cm}
$^*$ Service de Physique de l'\'Etat Condens\'e,\\
 Centre d'\'etudes de Saclay,\\
Orme des Merisiers,
91191 Gif-sur-Yvette {\sc cedex}, FRANCE\\
\vspace{0.4cm}
First version: June 2, 2000\\
{\em This version}\\
\today}
\maketitle

\begin{abstract}
It is commonly believed that the correlations between stock returns
increase in high volatility periods. We investigate how much of these
correlations can be explained within a simple non-Gaussian one-factor
description with {\it time independent} correlations. Using surrogate
data with the true market return as the dominant factor, we show that
most of these correlations, measured by a variety of different
indicators, can be accounted for. In particular, this one-factor model
can explain the level and asymmetry of empirical exceedance correlations.  
However, more subtle effects require an extension of the one factor model, 
where the variance and skewness of the residuals also depend on the market
return.
\end{abstract}

\section{Introduction}

Understanding the relationship between the statistics of individual
stock returns and that of the corresponding index is a major issue in
several finance problems such as risk management \cite{Elton}
or market micro-structure modeling. It is also crucial for building
optimized portfolios containing both index and stocks derivatives
\cite{Hull,Taleb}.  Although the index return is the (weighted) sum of
stock returns, it actually displays very different statistical
properties from what would result if the stock
returns were independent. In particular, the cumulants (that is, 
the volatility, the
skewness and the kurtosis) of the index distribution, which should be
suppressed by a power of the number of stocks $N$ for independent returns,
are still very large, even for $N=500$. The negative skewness of the index, 
in particular,
is actually larger than for individual stocks, and reflects a specific leverage
effect \cite{Leverage}. 

It is a common belief that cross-correlations between stocks actually
{\it fluctuate}\/ in time, and increase substantially in a period of
high market volatility. This has been discussed in many
papers -- see for example \cite{Corrhigh1}, with more recent discussions, including new 
indicators, in \cite{Longin,Ang1,Ang2,Allemands}. Furthermore, this increase 
is thought to be larger for large downward moves than for large upward moves. 
The dynamics of these correlations themselves, and their asymmetry,
should be estimated, leading to rather complex models \cite{Ang1,Bekk,lo,Beckaert}. 
The view of `moving' correlations 
has a direct consequence for risk management: the
risk for a given portfolio is seen as resulting from both volatility 
fluctuations and correlation fluctuations.

An alternative point of view is provided by factor models with a
fixed correlation structure. The simplest version contains a unique factor 
-- the market itself. In this case, the time fluctuations of the measured 
cross-correlations between stocks is, as we show below, directly related to the
fluctuations of the market volatility. The notion of ``correlation''
risk therefore reduces to market volatility risk, which considerably
simplifies the problem.  In this paper, we want to address to what
extent a {\it non-Gaussian}\/ one-factor model is able to capture the
essential features of stocks cross-correlations, in particular in
extreme market conditions. Our conclusion is that most of the extreme
risk correlations, measured by different indicators, 
are actually captured by this simple {\it fixed-correlation} model. 
This model is able to reproduce quantitatively the observed {\it exceedance
correlations} \cite{Longin} without invoking the idea of `regime switching' recently advocated
in this context in \cite{Ang1,Ang2}.

However, a more detailed analysis shows that a
refined model is needed to account for the dependence of the
conditional volatility and skewness of the residuals on the market
return.

\section{A non-Gaussian one-factor model}

We want to compare empirical measures of correlation with the
prediction of a {\it fixed-correlation}\/ model. However, for generic
non-Gaussian probability distributions of returns, there is no unique
way of building a multivariate process. A natural choice is to assume
that the return of every stock is the sum of random independent
(non-Gaussian) factors.  While a multivariate Gaussian process can
always be decomposed into independent factors, this is not true for
generic non-Gaussian distributions. The existence of such a
decomposition is thus part of the definition of our model.  

\paragraph{The model:} We will call
{\em market}\/ the dominant factor in this decomposition and write:
\begin{equation}
\label{eq:onefactor}
r_i(t)=\beta_i r_m(t) + \epsilon_i(t).
\end{equation}
The daily return is defined as $r_i(t)=S_i(t)/S_i(t-1)-1$, where
$S_i(t)$ is the value of the stock $i$ on day $t$. The return is thus
decomposed into a market part $r_m(t)$ and a residual part
$\epsilon_i(t)$.  In a generic factor model, the residuals
$\epsilon_i(t)$ are combinations of all the factors except the market
and are therefore independent of it.  The {\em one-factor model}\/
corresponds to the simple case where the $\epsilon_i(t)$ are also
independent of one another.

The market is defined as a weighted sum of the returns of all stocks.
The weights can be those of a market index such as the S\&P 500.  These
could also be the components of the eigenvector with the largest
eigenvalue of the stocks cross-correlation matrix \cite{CAPM}. We have
chosen to work simply with uniform weights, leading to the following definition:
\begin{equation}
r_m=\frac{1}{N}\sum_{i=1}^N r_i.
\label{eq:market}
\end{equation}
Had we chosen another weighting scheme for the definition of the market,
the theoretical results below would still hold exactly provided that we replace
averages over all stocks by the corresponding weighted averages. On our data set, the 
different weighted averages give essentially the same results.
The coefficients $\beta_i$'s are then given by:
\begin{equation}
\label{eq:beta}
\beta_i=\frac{\langle r_ir_m\rangle - \langle r_i \rangle \langle r_m
\rangle}{\langle r_m^2 \rangle -\langle r_m \rangle ^2},
\end{equation}
where the brackets refer to time averages.
This model is meaningful in the case where $\beta_i$ is constant or slowly 
varying in time. Eq.~(\ref{eq:market}) immediately
implies $(1/N) \sum_{i=1}^N\beta_i=1$.

An important qualitative assumption of this model is that although the
market is built from the fluctuations of the stocks, it is
a more fundamental quantity than the stocks themselves. Hence, one
cannot expect to explain the statistical properties of the
market from those of the stocks within this model.

\paragraph{Real data and surrogate data:} The data set we considered 
is composed of the daily returns of 450
U.S. equities among the most liquid ones from 1993 up to 1999. In
order to test the validity of a one-factor model, we also generated
surrogate data compatible with this model. Very importantly, the
one-factor model we consider is not based on Gaussian distributions,
but rather on fat-tailed distributions that match the empirical
observations for both the market and the stocks daily returns
\cite{levy-empirical}.

The procedure we used to generate the surrogate data is the following:
\begin{itemize}
\item Compute the $\beta_i$'s using Eq.~(\ref{eq:beta}) over the whole time
period \cite{beta}. These $\beta_i$'s are found to be rather 
narrowly distributed around $1$, with $(1/N) \sum_{i=1}^N\beta_i^2=1.05$. 
\item Compute the variance of the residuals
$\sigma_{\epsilon_i}^2=\sigma_i^2-\beta_i^2\sigma_m^2$.
On the dataset we used $\sigma_m$ was 0.91\% (per day) whereas
the rms $\sigma_{\epsilon_i}$ was 1.66\%. 
\item Generate
the residual $\epsilon_i(t)=\sigma_{\epsilon_i}u_i(t)$, where the $u_i(t)$
are independent random
variables of unit variance with a leptokurtic (fat tailed) distribution
--- we have chosen here a Student distribution with an exponent $\mu=4$:
\begin{equation}
P(u)=\frac{3}{(2+u^2)^{5/2}},
\end{equation}
which is known to represent adequately the empirical data \cite{levy-empirical}.
\item Compute the surrogate return as $r_i^{\mbox{\tiny
surr}}(t)=\beta_i r_m(t)+\epsilon_i(t)$, where $r_m(t)$ is the {\it
true}\/ market return at day $t$. 
\end{itemize}
Therefore, within this method, both the empirical and surrogate returns are based on the 
very same
realization of the market statistics. This allows us to compare meaningfully the results
of the surrogate model with real data, without further averaging. It also short-cuts 
the precise parameterization of
the distribution of market returns, in particular its correct negative skewness, which
turns out to be crucial.

\section{Conditioning on large returns}
\paragraph{Conditioning on absolute market return:}
We have first studied a measure of correlations
between stocks conditioned on an extreme market return. It is indeed 
commonly believed that cross-correlations between stocks increase in such
``high-volatility'' periods. A natural measure is given by the following coefficient:
\begin{equation}
\rho_>(\lambda)=\frac{\frac{1}{N^2}\sum_{i,j} \left ( \langle r_i r_j
\rangle_{>\lambda} - \langle r_i \rangle_{>\lambda} \langle
r_j \rangle_{>\lambda} \right )} {\frac{1}{N}\sum_i \left (
\langle r_i^2 \rangle_{>\lambda} - \langle
r_i\rangle_{>\lambda}^2 \right )},
\end{equation}
where the subscript $>\lambda$ indicates that the averaging is
restricted to market returns $r_m$ in absolute value larger than $\lambda$.
For $\lambda=0$ the conditioning disappears. Note that the quantity
$\rho_>$ is the average covariance divided by the average
variance, and therefore differs from the average correlation
coefficient. We have studied the latter quantity, and the
following conclusions remain valid in this case also.

\begin{figure}
\psfrag{xaxis}[ct][ct]{\small $\lambda$ (\%)}
\psfrag{yaxis}[cb][cb]{\small $\rho_>(\lambda)$}
\psfrag{legend1}[l][l]{\small Empirical data}
\psfrag{legend2}[l][l]{\small One factor model}
\centerline{\epsfig{file=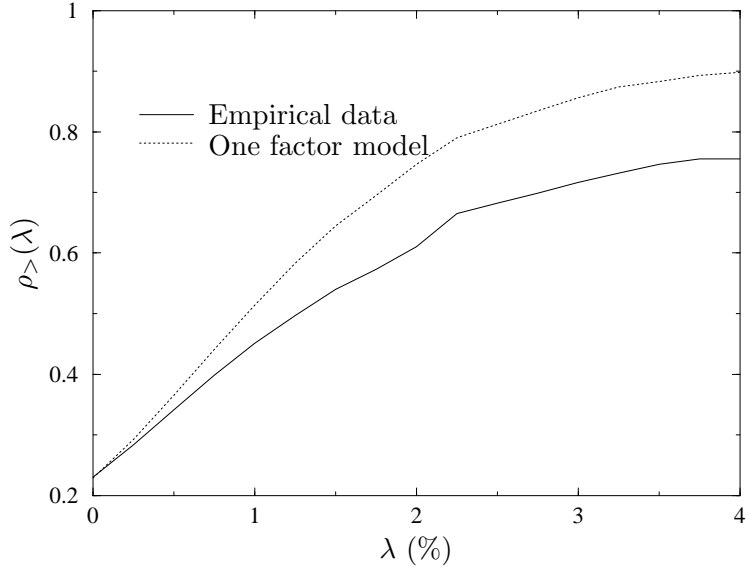,height=0.35\textheight}}
\caption{Correlation measure $\rho_>(\lambda)$ conditional to the
absolute market return to be larger than $\lambda$, both 
for the empirical data and the one-factor model. Note that both show a 
similar apparent increase of correlations with $\lambda$. This effect is actually 
overestimated by the one-factor model with fixed residual volatilities.
$\lambda$ is in percents.}
\label{fig:condcorr}
\end{figure}

In a first approximation, the distribution of individual stocks returns can be
taken to be symmetrical, leading $\langle r_i \rangle_{>\lambda} \simeq 0$.
The above equation can therefore be transformed into:
\begin{equation}
\label{eq:rho}
\rho_>(\lambda)
\simeq \frac{\sigma_m^2(\lambda)}{\frac{1}{N}\sum_{i=1}^N\sigma_{i}^2(\lambda)},
\end{equation} 
where $\sigma_m^2(\lambda)$ is the market volatility conditioned to market
returns in absolute value larger than $\lambda$, and 
$\sigma_i^2(\lambda)=\langle r_i^2 \rangle_{>\lambda} -
\langle r_i\rangle_{>\lambda}^2$. In the context of a one-factor
model, we therefore obtain:
\begin{equation}
\label{eq:rhoonefactor}
\rho_>(\lambda) = \frac{\sigma_m^2(\lambda)}{\left(\frac{1}{N}\sum_{i=1}^N\beta_i^2\right)
\sigma_m^2(\lambda) + \frac{1}{N}\sum_{i=1}^N\sigma_{\epsilon_i}^2}.
\end{equation} 
The residual volatilities $\sigma_{\epsilon_i}^2$ are independent of $r_m$ 
and therefore of $\lambda$ whereas $\sigma_m^2(\lambda)$ is obviously an increasing 
function of $\lambda$. Hence the coefficient $\rho_>(\lambda)$ is an increasing function of $\lambda$.
The one-factor model therefore predicts an increase of the
correlations (as measured by $\rho_>(\lambda)$) in high volatility periods. 
This conclusion is quite general, it holds in particular for any factor model, even with
Gaussian statistics. Therefore, the very fact of conditioning the correlation on 
large market returns leads to an increase of the measured correlation. 
A similar discussion in the context of Gaussian models can be found in \cite{Longin}.

More precisely, we can now compare the coefficient $\rho_>(\lambda)$ measured
empirically to one obtained within the one-factor model defined above. 
This is presented in Fig.~\ref{fig:condcorr}. Interestingly, the surrogate and empirical
correlations are similar, displaying qualitatively
the same increase of the cross-correlation when conditioned to large
market returns. This shows that a one-factor model does indeed account
quantitatively for the apparent increase of cross-correlations in high volatility
periods. 

The one-factor model actually even {\it overestimates} the correlations
for large $\lambda$. This overestimation can be understood
qualitatively as a result of a positive correlation between the
amplitude of the market return $|r_m|$ and the residual volatilities
$\sigma_{\epsilon_i}$, which we discuss in more details in Section 4 below (see in
particular Fig.~5). 
For large values of $\lambda$, $\sigma_{\epsilon_i}$ is found to be larger than its
average value.  From Eq.~(\ref{eq:rhoonefactor}), this lowers the
correlation $\rho_>(\lambda)$ as compared to the simplest one-factor
model where the volatility fluctuations of the residuals are neglected.

\begin{figure}
\psfrag{xlabel}[ct][ct]{\small $r_m$ (\%)} 
\psfrag{ylabel}[cb][cb]{\small $P(\mbox{sign}(r_i)=\mbox{sign}(r_m))$} 
\psfrag{legend1}[l][l]{\small Empirical data} 
\psfrag{legend2}[l][l]{\small One factor model}
\centerline{\epsfig{file=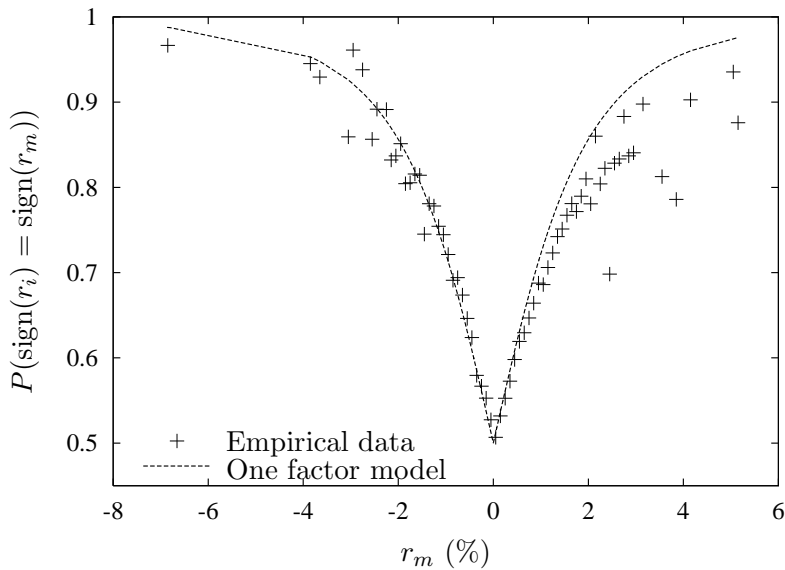,height=0.35\textheight}}
\caption{Conditional probability that a stock has the same sign as the market return
as a function of the market return; $r_m$ is in percent. Each cross represents
the empirical probability using 4\% of the days centered around a given market return. 
The dotted line is the prediction of the non-Gaussian one-factor model.}
\label{fig:fracreturncondindex}
\end{figure}

\paragraph{Conditional fraction of positive/negative returns:}
Another quantity of interest is the fraction of stocks returns having
the same sign as the market return, as a function of the market
return itself. 
The empirical results are shown on Fig.~\ref{fig:fracreturncondindex}. 
We observe that for the largest
returns, $90\%$ of the stocks have the same return sign as that of the
market. Therefore, the sign of the market appears to have a very strong influence
on the sign of individual stock returns.

This fraction can be calculated exactly within the one-factor
model. Focusing on positive market return (the case of negative returns
can be treated similarly), a stock return $r_i$ is positive whenever
$\epsilon_i>-\beta_i r_m$. Therefore the average fraction $f(t)$ of
stocks having a positive return for a given market return $r_m(t)$ is
\begin{equation}
f(t)=\frac{1}{N}\sum_{i=1}^N {\cal P}_< \left (\frac{\beta_i
r_m(t)}{\sigma_{\epsilon_i}} \right ),
\end{equation}
where ${\cal P}_<$ is the cumulative normalized 
distribution of the residual (chosen here to be a Student 
distribution with a exponent $\mu=4$). $f(t)$ is also plotted on
Fig.~\ref{fig:fracreturncondindex} and fits well the empirical
results. The theoretical estimate slightly overestimates the fraction $f(t)$ 
for positive market returns. As explained above, the correlations between
$\sigma_{\epsilon_i}$ and $|r_m|$ do lower $f(t)$ as needed for the positive
side. However, the corresponding fraction for the negative side would then be
underestimated.

\paragraph{Conditioning on large individual stock returns -- quantile correlations and
exceedance correlations:}

\begin{figure}
\psfrag{xlabel}[ct][ct]{\small $q$}
\psfrag{ylabel}[cb][cb]{\small $\rho(q)$}
\psfrag{legend1}[l][l]{\small Empirical data}
\psfrag{legend2}[l][l]{\small Surrogate data}
\centerline{\epsfig{file=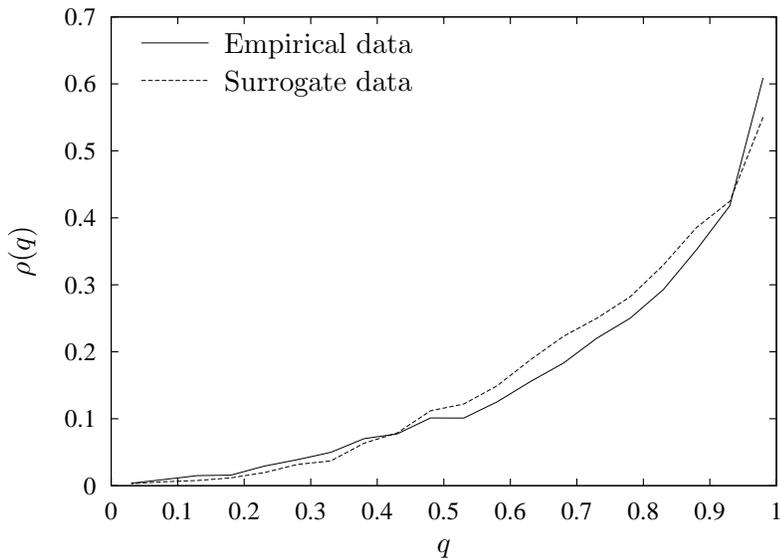,height=0.35\textheight}}
\caption{Correlation between stocks for joint extreme moves, $\rho(q)$, as a 
function of the quantile value $q$, both for real data and 
the surrogate one-factor model.}
\label{fig:quantcorr}
\end{figure}

Since the volatility of the residuals is two times larger than the
volatility of the market, the conditioning by extreme market events
does not necessarily select extreme individual stock moves.  The
quantities studied in the previous section, namely return correlations
and sign correlations, are therefore more related to the central part
of the stocks distribution rather than to their extreme tails.  We now
study more specifically how extreme stock returns are correlated
between themselves. A first possibility is to study {\it quantile correlations},
that we define as:
\begin{equation}
\rho(q)=\frac{\frac{1}{N^2}\sum_{i,j} \left ( \langle r_i r_j
\rangle_{q} - \langle r_i \rangle_{q} \langle
r_j \rangle_{q} \right )} {\frac{1}{N}\sum_i \left (
\langle r_i^2 \rangle_{q} - \langle
r_i\rangle_{q}^2 \right )},
\end{equation}
where the subscript $q$ indicates that we only retain in the average
days such that {\it both}\/ $|r_i|$ and $|r_j|$ take their $q-$quantile
value, within a certain tolerance level (this tolerance is taken to be $4\%$
of the total interval for each quantile). In the limit $q \to 1$, this
selects extremes days for both stocks $i$ and $j$ simultaneously. The
empirical results for $\rho(q)$ are compared with the prediction of
the one-factor model in Fig. \ref{fig:quantcorr}. The agreement is
again very good, though the one-factor model still slightly
overestimates the true correlations in the extremes.

\begin{figure}
\psfrag{xaxis}[ct][ct]{\small $\theta$}
\psfrag{yaxis}[cb][cb]{\small $\rho^{\pm}(\pm \theta)$}
\psfrag{legend1}[l][l]{\small Empirical data}
\psfrag{legend2}[l][l]{\small Surrogate data}
\centerline{\epsfig{file=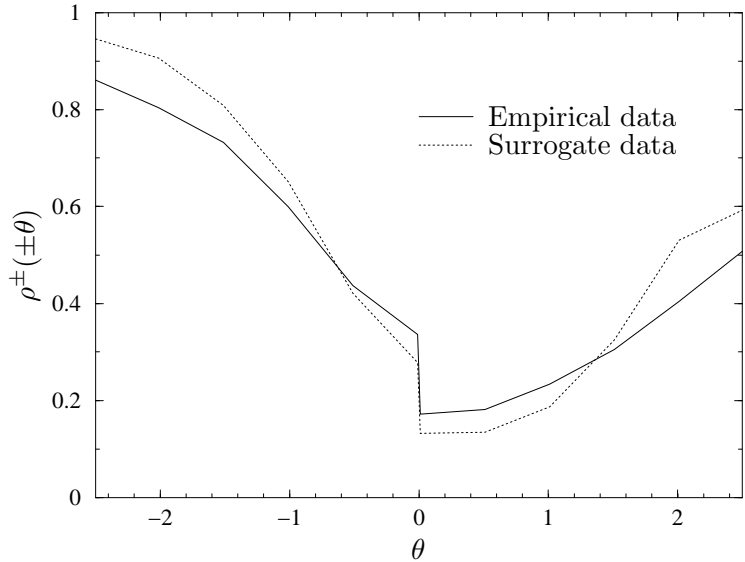,height=0.35\textheight}}
\caption{Average exceedance correlation functions between stocks 
as a function of the level parameter $\theta$, both for real data and 
the surrogate one-factor model. We have shown $\rho_{ij}^+(\theta)$
for positive $\theta$ and $\rho_{ij}^-(-\theta)$ for negative $\theta$.
Note that this quantity {\it grows} with $|\theta|$ and is strongly asymmetric.}
\label{fig:exceedcorr}
\end{figure}

Another interesting quantity that has been much studied in the econometric
literature recently, is the so-called exceedance correlation function, introduced in
\cite{Longin}. One first defines normalized centered returns $\tilde r_i$ with zero
mean and unit variance. The positive exceedance correlation between $i$ and $j$ is 
defined as:
\be
\rho_{ij}^+(\theta)=\frac{\langle \tilde r_i \tilde r_j \rangle_{>\theta}
- \langle \tilde r_i \rangle_{>\theta} \langle \tilde r_j \rangle_{>\theta}}
{\sqrt{(\langle \tilde r_i^2 \rangle_{>\theta}
- \langle \tilde r_i \rangle_{>\theta}^2)(\langle \tilde r_j^2 \rangle_{>\theta}
- \langle \tilde r_j \rangle_{>\theta}^2)}},
\ee
where the subscript $> \theta$ means that both normalized returns are larger than 
$\theta$. Large $\theta$'s correspond to extreme correlations. The negative exceedance
correlation $\rho_{ij}^-(-\theta)$ is defined similarly, the conditioning being now 
on returns smaller than $-\theta$. Fig.~\ref{fig:exceedcorr} shows the exceedance
correlation function, averaged over the pairs $i$ and $j$, both for real data and for the
surrogate one-factor model data. As in previous papers, we have shown $\rho_{ij}^+(\theta)$
for positive $\theta$ and $\rho_{ij}^-(-\theta)$ for negative $\theta$. As in previous studies
\cite{Longin,Ang1,Ang2},
we find that $\rho^{\pm}(\pm \theta)$ {\it grows} with $|\theta|$ and is larger for 
large negative moves than for large positive moves. This is in strong contrast with the
prediction of a Gaussian model, which gives a symmetric tent-shaped graph that goes to
zero for large $|\theta|$. Note however that previous studies have focused on fixed pairs
of assets $i$ and $j$ (for example a few pairs of international markets). The result
of Fig.~\ref{fig:exceedcorr} is interesting since it reveals a systematic effect over
all pairs of a pool of 450 stocks.  

Several models have been considered to explain the observed results \cite{Ang1,Ang2}.
Simple GARCH or Jump models cannot account for the shape of the exceedance correlations.
Qualitatively similar graphs can however be reproduced within a rather sophisticated 
`regime switching' model, where the two assets switch between a positive, low volatility 
trend with small cross-correlations
and a negative, high volatility trend with large cross-correlations. Note that by construction,
this `regime switching' model induces a strong skew in the `index' (i.e. the average between
the two assets). Fig.~\ref{fig:exceedcorr} however clearly shows that a {\it fixed} correlation
non Gaussian one-factor model is enough to explain quantitatively the level and asymmetry
of the exceedance correlation function. In particular, the asymmetry is induced by
the large negative skewness in the distribution of index returns, and the growth of the
exceedance correlation with $|\theta|$ is related to distribution tails fatter than exponential
(in our case, these tails are indeed power-laws).

\section{Conditional statistics of the residuals}

We conclude from the above results that the observed fluctuations of
the stock cross-correlations are mainly a consequence of the
volatility fluctuations and skewness of 
the market return, and that a non Gaussian
one-factor model does reproduce satisfactorily most of the observed
effects. However, some small systematic discrepancies appear, and call
for an extension of the one-factor model.  The most obvious effect not
captured by a one-factor model is the recently discovered `ensemble' skewness
in the daily distribution of stock returns, as discussed by Lillo
and Mantegna \cite{Lillo}. More precisely, they have shown that the
histogram of all the stocks returns {\it for a given day}\/ displays on
average a positive skewness when the market return is
positive, and a negative skewness when the market return is
negative. The amplitude of this skewness furthermore grows with the
absolute value of the market return. Note that this skewness is {\it not}
related to the possible non zero skewness of individual stocks that has 
been recently discussed in several papers in relation with extended CAPM models 
\cite{Beckaert,Harvey}.

\begin{figure}
\psfrag{xlabel}[ct][ct]{$|r_m|$ (\%)}
\psfrag{ylabel}[cb][cb]{$\Sigma$ (\%)}
\psfrag{legend1}[l][l]{\small Empirical data}
\psfrag{legend2}[l][l]{\small Linear fit: $\rho=0.54$}
\centerline{\epsfig{file=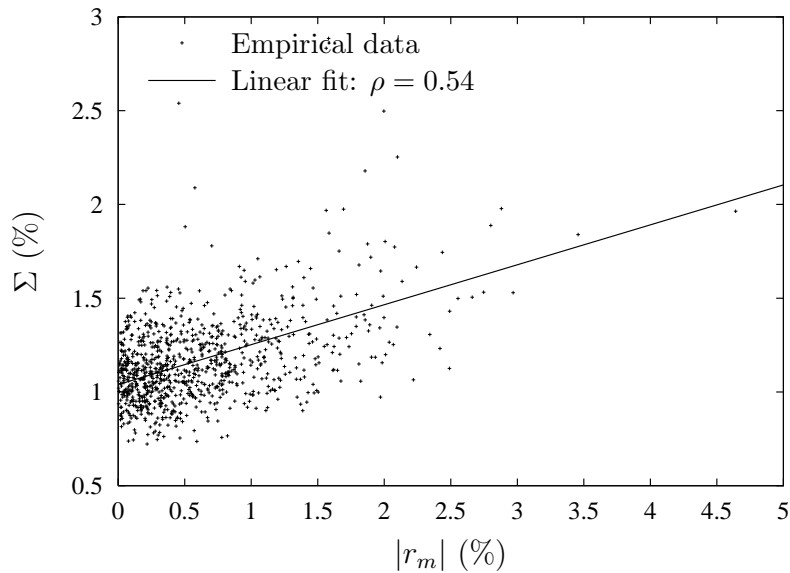,height=0.35\textheight}}
\caption{Daily residual `volatility' $\Sigma$ (Eq.~(\ref{bigsig})) averaged over the different 
stocks for a given day as a function of the market return for the same day. 
$\Sigma$ is in percent.}\label{fig:condvariance}
\end{figure}

Clearly, this `ensemble' skewness that depends on the market return
cannot be explained by the above one-factor model where the residuals have
a time independent zero skewness. The one-factor
model is certainly an oversimplification of the reality: although the
market captures the largest part of the correlation between stocks,
industrial sectors are also important, as can be seen from a
diagonalization of the correlation matrix \cite{Mantegna}.  Large moves of the market
can be dominated by extreme moves of a single sector, while the other
sectors are relatively unaffected. This effect does induce some skewness in
the fixed-day histogram of stock returns distribution.

A way to account for this effect is to allow the distribution of the
residual $\epsilon_i(t)$ to depend on the market return $r_m(t)$. In
order to test this idea, we have studied directly some moments of the
distribution of the residuals {\it for a given day}\/ as a function of
the market return that particular day. We have studied the following
quantities:
\begin{equation}\label{bigsig}
\Sigma = [|\epsilon_i -[\epsilon_i] |], 
\end{equation}
\begin{equation}\label{bigS}
{\cal S}=\frac{[ \epsilon_i ] -\mbox{Med}(\epsilon_i)}{\Sigma},
\end{equation}
\begin{equation}\label{bigK}
{\cal K}=\frac{[\epsilon_i^2] - [\epsilon_i]^2}{\Sigma^2}, 
\end{equation}
where the square brackets $[...]$ means that we average over the
different stocks for a given day and $\mbox{Med}$ selects the median
value of $\epsilon_i$.  These three quantities should be thought as {\em
robust}\/ alternatives to the standard variance, skewness and kurtosis,
which are based on higher moments of the distribution.

\begin{figure}
\psfrag{xlabel}[ct][ct]{$r_m$ (\%)}
\psfrag{ylabel}[cb][cb]{${\cal S}$}
\psfrag{legend1}[l][l]{\small Empirical data}
\psfrag{legend2}[l][l]{\small Linear fit: $\rho=0.44$}
\centerline{\epsfig{file=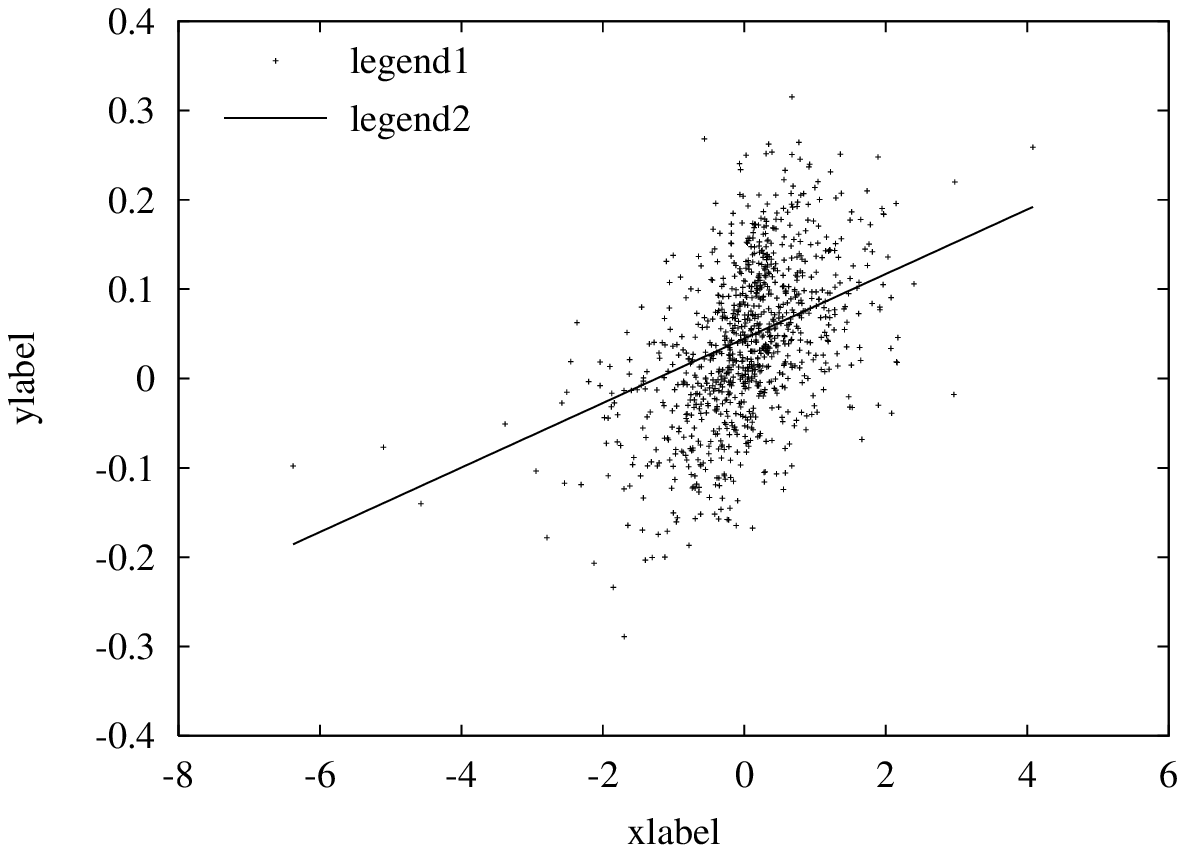,height=0.35\textheight}}
\caption{Daily residual `skewness' $\cal S$ averaged over the different 
stocks for a given day as a function of the market return for the same day. 
Note that the skewness is computed using low moments of the 
distribution to reduce the 
measurement noise and does not correspond to the usual definition (see Eq.~(\ref{bigS})).}
\label{fig:condskew}
\end{figure}

The quantity $\Sigma$ measures the `volatility' of the residuals and
is shown in Fig.~\ref{fig:condvariance} as a function of $|r_m(t)|$. A
linear regression is also shown for comparison. It is clear
that there is a positive correlation between the market volatility and
the volatility of the residuals, not captured by the simplest
one-factor model. As explained above, this effect actually allows one
to account quantitatively for the systematic overestimation of the observed
correlations.

In order to confirm the skewness effect of Lillo and Mantegna, we have
then studied the quantity ${\cal S}$.  This quantity is positive if the
distribution is positively skewed.  Fig.~\ref{fig:condskew} shows a
scatter plot of ${\cal S}$ as a function of $r_m$
\cite{NoteSkew}. Again these two quantities are positively correlated,
as emphasized by Lillo and Mantegna (although their analysis is different
from ours).

\begin{figure}
\psfrag{xlabel}[ct][ct]{$|r_m|$ (\%)}
\psfrag{ylabel}[cb][cb]{${\cal K}$}
\psfrag{legend1}[r][r]{\small Empirical data}
\psfrag{legend2}[r][r]{\small Linear fit: $\rho=-0.13$}
\centerline{\epsfig{file=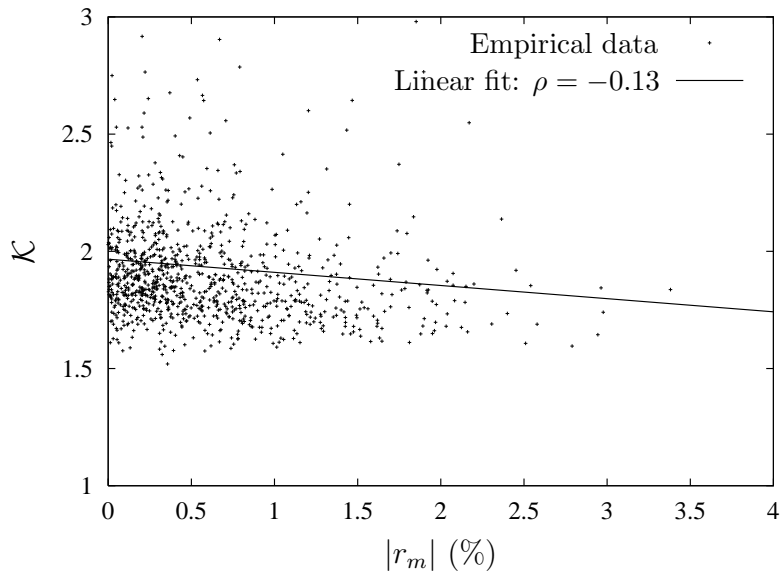,height=0.35\textheight}}
\caption{Daily residual `kurtosis' averaged over the different stocks 
for a given day as a function of the market return for the same day. 
Again, the kurtosis is computed using low moments of the distribution 
to reduce the measurement noise and does not correspond to the usual 
definition (see Eq.~(\ref{bigK})).}
\label{fig:condkurt}
\end{figure}

Therefore, both the volatility and the skew of the residuals are quite
strongly correlated with the market return. One could wonder if higher
moments of the distribution are also sensitive to the value of $r_m$.
We have therefore studied the quantity ${\cal K}$ as one possible
refined measure of the shape of the distribution of residuals. This is
shown in Fig.~\ref{fig:condkurt} and reveals a much weaker dependence
than the previous two quantities.

\section{Conclusion}

We have thus shown that the apparent increase of correlation between stock 
returns in extreme conditions can be satisfactorily explained within a
{\it static} one-factor model which accounts for fat-tail effects. In this model,
conditioning on a high observed volatility naturally leads to an increase 
of the apparent correlations. The much discussed exceedance correlations can 
also be reproduced quantitatively and reflects both the non-Gaussian nature
of the fluctuations and the negative skewness of the index, and {\it not} the fact
that correlations themselves are time dependent.

This one-factor model is however only an 
approximation to the true correlations, and more subtle effects (such
as the Lillo-Mantegna `ensemble' skewness) require an
extension of the one factor model, where the variance and skewness of
the residuals themselves depend on the market return.

\paragraph{Acknowledgments:}  We wish to thank M. Meyer and J. Miller for 
many useful discussions.


\begin{thebibliography}{99}
\bibitem{beta}  We do not think that computing
$\beta_i$ and $\sigma_{\epsilon_i}$ ``in sample'' has any consequence
on the presented results.
\bibitem{NoteSkew} Due to the discreteness of quoted prices, 
a substantial fraction of the stocks closes at the same price
two days in a row, producing a zero return. This large number of exactly
zero returns often makes the median return to be zero which in turns
induces an uninteresting systematic effect in the quantity $\cal S$.
To get rid of this effect we have added to each return of random number
uniformly distributed between -0.25\% and 0.25\%.
\end{thebibliography}
\end{document}